# Phase Aberration Correction with Adaptive Coherence-Weighted Point Spread Function Restoration Filtering Technique

Wei-Hsiang Shen
*Department of Electrical Engineering*
*National Tsing Hua University*
Hsinchu, Taiwan
whshen@gapp.nthu.edu.tw

Yu-An Lin
*Department of Electrical Engineering*
*National Tsing Hua University*
Hsinchu, Taiwan

Pai-Chi Li
*Department of Electrical Engineering*
*Insitute of Biomedical Electronics and Bioinformatics*
*Institue of Medical Device and Imaging*
*Graduate School of Advanced Technology*
*National Taiwan University*
Taipei, Taiwan

Meng-Lin Li
*Department of Electrical Engineering*
*Institute of Photonics Technologies*
*Brain Research Center*
*National Tsing Hua University*
Hsinchu, Taiwan
mlli@ee.nthu.edu.tw

*Abstract*—Phase aberration is an inherent side effect of ultrasound imaging due to the speed of sound inhomogeneity nature of human tissues, resulting in focusing error and reduced image contrast. This work introduces a phase aberration correction technique by leveraging a point spread function (PSF) restoration filter. A convolutional neural network (CNN) is used to estimate phase-aberrated PSFs and design the restoration filter. In addition, we incorporate coherence index weighting, derived from the restoration filtering, to further suppress sidelobe energy. Evaluation using Field II-simulated phantoms showed clearer cyst borders and reduced sidelobe energy leakage after PSF restoration and filter-derived coherence weighting, leading to improvement in image contrast and quality.

## I. Introduction

Phase aberration is an inevitable side effect in ultrasound imaging due to the speed of sound inhomogeneity nature of the imaged human tissue. This inhomogeneity contradicts the constant speed of sound assumption during ultrasound beamforming [1]–[3]. It distorts the amplitude and phase of the acoustic signal and degrades image contrast and quality, reducing the diagnosis value of ultrasound images. Such degradation is especially pronounced in tissues where fat accumulates, such as the breast and liver, due to the higher inhomogeneity of speed of sound within these tissue [4].

Phase aberration in ultrasound imaging has been a persistent challenge. Some attempted to correct phase aberration by estimating the aberration profile through the correlation of channel data [1], [2]. Others extended the concept by calculating coherence factor to further reduce focusing error and suppress sidelobe induced by phase aberration [5]. More recent studies have employed deep learning techniques to estimate the aberration profile or to directly correct phase aberration on log-compressed images [6], [7]. However, these approaches either use channel data or need to cache channel data for refocusing, which is memory inefficient and thus impractical to implement on most commercial ultrasound scanners.

In our previous work, we approached the phase aberration correction problem using a point spread function (PSF) restoration filtering technique [8]. This filter is designed by an estimated phase aberrated PSF predicted by a neural network and the ideal aberration-free PSF. This filter can be applied on beamformed baseband data (i.e., IQ data) to correct the degradation caused by phase aberration. Our results showed improvement in objective metrics such as contrast ratio (CR) and contrast to noise ratio (CNR).

In this work, we improve our previously proposed PSF restoration filtering technique by deriving an adaptive coherence index from the filtering procedure. High coherence indicates high mainlobe energy, whereas low coherence indicates high sidelobe energy. Therefore, this index can be used as a weighting factor to further suppress the sidelobe energy caused by phase aberration, improving imaging contrast and quality.

## II. Materials and Methods

An ultrasound imaging system can be assumed as a locally linear space-invariant system. That is, the PSF within a local imaging area remains consistent across all spatial locations. Therefore, if a filter could restore one phase-aberrated PSF into an ideal aberration-free PSF, an ideally focused image can then be reconstructed by the filter. Based on the assumption, we can design such PSF resrestoration filter for phase aberration correction.

### A. PSF Restoration Filter

We design an adaptive inverse filter that can restore a phase-aberrated PSF into its ideal aberration-free form. The restoration filter K can be described as,

$$PSF_{ideal} = PSF_{aberrated} * K \quad (1)$$

where $PSF_{ideal}$ is the ideal aberration-free PSF, $PSF_{aberrated}$ is the phase aberrated PSF, K is the reshaping filter, and $*$ is the convolution operator.

An ultrasound imaging system can be assumed as locally linear space-invariant and thus can be modeled in the convolution form as [8],

$$R_{aberrated} = S * PSF_{aberrated} \quad (2)$$

where $R_{aberrated}$ is the observed (thus phase-aberrated) baseband data, and S is the tissue scatterer distribution. In addition, we can define the corresponding ideal aberration-free system as,

$$R_{ideal} = S * PSF_{ideal} \quad (3)$$

where $R_{ideal}$ is the aberration-free baseband data. Therefore, the restoration filter K can be used on the phase-aberrated baseband data and reconstructs the aberration-free data because of the associative property of the convolution operator.

$$R_{aberrated} * K = S * PSF_{aberrated} * K \quad (4)$$
$$= S * PSF_{ideal} \quad (5)$$
$$= R_{ideal} \quad (6)$$

With equation 1, we can solve for K using the analytical solution from L2 optimization [9], [10]. Since $PSF_{ideal}$ can be precomputed given the imaging parameters, the design of the PSF reshaping filter relies on accurately estimating the underlying phase-aberrated PSF $PSF_{aberrated}$.

*B. Phase Aberrated PSF Esimation*

We employ a convolutional neural network (CNN) to estimate the underlying phase-aberrated PSF $PSF_{aberrated}$. This CNN takes a patch of beamformed radio frequency (RF) speckle patch as input and infers the underlying phase-aberrated PSF of the given patch. In this work, we utilize a U-Net3+ to estimate the phase-aberrated PSF [12].

The network is trained with synthetic data generated in Field II [13]. We simulate phase-aberrated PSFs following the near field phase screen model [1], [2]. This model assumes that there is a phase screen in front of the ultrasound transducer that applies time delay (so alters the phase) to the transmitted and received signals on each channel.

These synthetic phase aberrated PSFs are then convolved with Gaussian distributed scatterers to generate homogeneous speckle patches. The RF speckle patches and their corresponding RF PSFs together form the training pairs.

*C. Coherence-Weighted Filtering*

In addition to the PSF restoration filter, we can derive coherence index weighting from the filtering procedure to further suppress sidelobe and thus improve image contrast. At every spatial location, the coherence index is calculated using the sliding filter kernel during the convolution filtering.

The coherence index is defined as the ratio of the low-frequency energy to the total energy in the sliding point-wise product between the data and filter during convolution, which indicates the concentration of low-frequency energy [5], [11].

For signals in the mainlobe region, the sliding point-wise product of the filter and baseband data would have high coherence, whereas for sidelobe signals, the sliding point-wise product would have low coherence, as shown in Figure 1. Therefore, the derived coherence index can be effectively used as a weighting factor to suppress sidelobe energy of the filtered result.

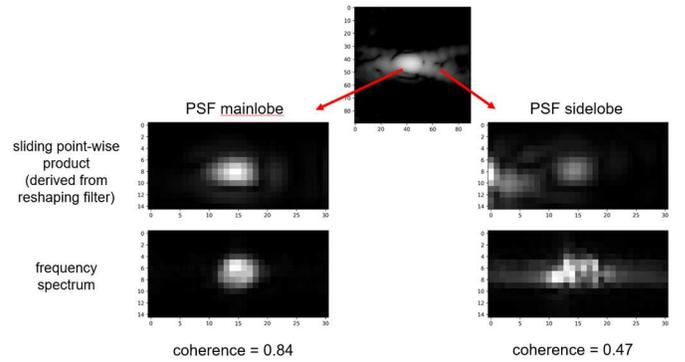

Fig. 1. An example of the derived coherence index at a mainlobe region and sidelobe region. Mainlobe region would have high coherence, whereas sidelobe region would have low coherence. Therefore, the derived coherence index can be used as a weighting factor to suppress sidelobe energy and improve image contrast.

III. RESULTS

We used a phase aberrated phantom simulated in Field II to evaluate the efficacy of the proposed technique, as shown in Figure 2. After the PSF restoration filter, the cyst's borders are clearer, with reduced sidelobe energy leakage. Moreover, after the coherence weighting technique, the sidelobe of the point target is effectively suppressed, leading to even more distinct borders for both the cyst and phantom boundary.

In addition, objective metrics such as contrast ratio (CR), contrast to noise ratio (CNR), and generalized contrast to noise ratio (gCNR) [14] are used to evaluate the performance gain of the proposed technique, as shown in Table 1. CR and gCNR have shown improvement after both PSF restoration and coherence weighting. However, there is a decline in CNR after coherence weighting. This is because the weighting alters the speckle pattern and thus increases speckle noise, causing decrease in CNR.

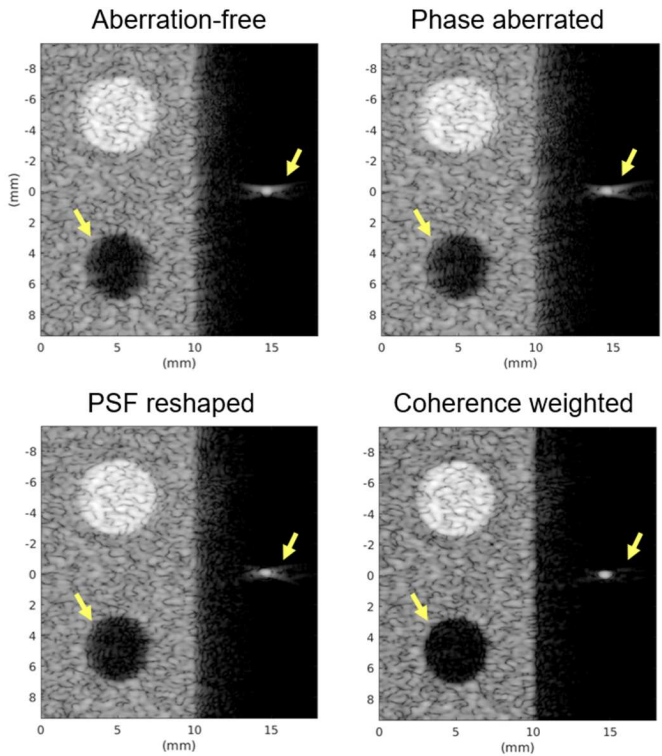

Fig. 2. Results of the proposed PSF restoration filter and coherence weighting technique. We can see clearer borders after PSF restoration (i.e., PSF reshaped), and effective sidelobe suppression of the point target after coherence weighting.

TABLE I. OBJECTIVE METRICS FOR PSF RESHPAING FILTER

|  | CR (dB) | CNR (dB) | gCNR |
|---|---|---|---|
| Aberration-free (ideal) | 20.00 | 4.39 | 0.767 |
| with phase aberration | 19.35 | 4.33 | 0.715 |
| + PSF restoration filter | 20.12 | 4.44 | 0.760 |
| + coherence weighting | 21.27 | 4.32 | 0.793 |

## IV. CONCLUSIONS

In this study, we improve the PSF restoration filtering technique by integrating the coherence index weighting technique to further suppressing sidelobe energy. Results showed improvement in image contrast and quality after the PSF restoration filter and the coherence weighting technique. Our proposed technique provides a solution for phase aberration correction on baseband data.

ACKNOWLEDGMENT

This research is supported by National Science and Technology Council, Taiwan (MOST 110-2221-E-007-011-MY3)